\documentclass[nocrop]{bioinfo}

\usepackage{nameref}

\newcommand{\bdm}{BioDynaMo}
\newcommand{\systemA}{System A (see Supplementary File S1 Table~4)}
\newcommand{\systemB}{System B (see Supplementary File S1 Table~4)}
\newcommand{\systemC}{System C (see Supplementary File S1 Table~4)}

\copyrightyear{2015} \pubyear{2015}

\access{Advance Access Publication Date: Day Month Year}
\appnotes{Original paper}

\begin{document}
\firstpage{1}

\subtitle{Systems biology}

\title[\bdm{}]{BioDynaMo: a general platform for scalable agent-based simulation}
\author[Breitwieser \textit{et~al}.]{
Lukas Breitwieser\,$^{\text{\sfb 1,2,}*}$,
Ahmad Hesam\,$^{\text{\sfb 1,3,}*}$,
Jean de Montigny\,$^{\text{\sfb 1}}$,
Vasileios Vavourakis\,$^{\text{\sfb 4,5}}$,
Alexandros Iosif\,$^{\text{\sfb 4}}$,
Jack Jennings\,$^{\text{\sfb 6}}$,
Marcus Kaiser\,$^{\text{\sfb 6,7,8}}$,
Marco Manca\,$^{\text{\sfb 9}}$,
Alberto Di Meglio\,$^{\text{\sfb 1}}$,
Zaid Al-Ars\,$^{\text{\sfb 3}}$,
Fons Rademakers\,$^{\text{\sfb 1}}$,
Onur Mutlu\,$^{\text{\sfb 2,}*}$,
Roman Bauer$^{\text{\sfb 10,}*}$ 
}
\address{
$^{\text{\sf 1}}$ CERN openlab, CERN, European Organization for Nuclear Research, Geneva, Switzerland
\\
$^{\text{\sf 2}}$ ETH Zurich, Swiss Federal Institute of Technology in Zurich, Zurich, Switzerland
\\
$^{\text{\sf 3}}$ Delft University of Technology, Delft, the Netherlands
\\
$^{\text{\sf 4}}$ Department of Mechanical \& Manufacturing Engineering, University of Cyprus, Nicosia, Cyprus
\\
$^{\text{\sf 5}}$ Department of Medical Physics \& Biomedical Engineering, University College London, London, UK
\\
$^{\text{\sf 6}}$ School of Computing, Newcastle University, Newcastle upon Tyne, UK
\\
$^{\text{\sf 7}}$ Department of Functional Neurosurgery, Ruijin Hospital, Shanghai Jiao Tong University School of Medicine, Shanghai, China
\\
$^{\text{\sf 8}}$ Precision Imaging Beacon, School of Medicine, University of Nottingham, Nottingham, NG7 2UH
\\
$^{\text{\sf 9}}$ SCimPulse Foundation, Geleen, Netherlands
\\
$^{\text{\sf 10}}$ Department of Computer Science, University of Surrey, Guildford, UK 
}

\corresp{$^\ast$To whom correspondence should be addressed.}

\history{Received on XXXXX; revised on XXXXX; accepted on XXXXX}

\editor{Associate Editor: XXXXXXX}

\abstract {
  \textbf{Motivation:}
Agent-based modeling is an indispensable tool for studying complex biological systems.
However, existing simulators do not always take full advantage of modern hardware and often have a field-specific software design.\\
\textbf{Results:}
We present a novel simulation platform called \bdm{} that alleviates both of these problems. \bdm{} features a general-purpose and high-performance simulation engine. 
We demonstrate that \bdm{} can be used to simulate use cases in: neuroscience, oncology, and epidemiology.
For each use case we validate our findings with experimental data or an analytical solution.
Our performance results show that \bdm{} performs up to three orders of magnitude faster than the state-of-the-art baseline.
This improvement makes it feasible to simulate each use case with one billion agents on a single server, showcasing the potential \bdm{} has for computational biology research.\\
\textbf{Availability:} \bdm{} is an open-source project under the Apache 2.0 license and is available at \href{www.biodynamo.org}{www.biodynamo.org}.
Instructions to reproduce the results are available in supplementary information.\\
\textbf{Contact:} \normalsize \href{lukas.breitwieser@inf.ethz.ch}{lukas.breitwieser@inf.ethz.ch}, \href{a.s.hesam@tudelft.nl}{a.s.hesam@tudelft.nl}, \href{omutlu@ethz.ch}{omutlu@ethz.ch}, \href{r.bauer@surrey.ac.uk}{r.bauer@surrey.ac.uk}\\
\textbf{Supplementary information:} Available at \href{https://doi.org/10.5281/zenodo.4501515}{https://doi.org/10.5281/zenodo.4501515}.
 }

\maketitle

\section{Introduction}

Agent-based simulation is a powerful tool assisting life scientists in
  better understanding complex biological systems.
In silico simulation is an inexpensive and efficient way to rapidly test hypotheses
  about the (patho)physiology of cellular populations, tissues, organs, or
  entire organisms \citep{Yankeelov2016, ji2017mathematical}.

However, the effectiveness of such computer simulations for scientific research
  is often limited, mainly because of two reasons.
First, after the slowing down of Moore's law \citep{moores-law} and Dennard
  scaling \citep{dennard_design_1974}, hardware has become increasingly parallel
  and heterogeneous.
Most simulators do not take full advantage of these hardware enhancements.
The resulting limited computational power forces life scientists to
  compromise either on the resolution of the model or on simulation size \citep{thorne2007abm}.
Second, existing simulators have often been developed with a specific use case
  in mind.
This makes it challenging to implement the desired model, even if it deviates only
  slightly from its original purpose.

To help researchers tackle these two major challenges, 
we propose a novel open-source platform for biology dynamics modeling, \bdm{}.
We alleviate both of these problems by emphasizing performance and
  modularity.
\bdm{} features a high-performance simulation engine which is fully
  parallelized and able to offload computation to hardware accelerators.
The software comprises a set of fundamental biological functions, and a
  flexible design that adapts to specific user requirements.
Currently, \bdm{} implements the biological model presented in \cite{ZublerDouglas2009framework},
  but this model can easily be extended, modified, or replaced.
Hence, \bdm{} is well-suited for simulating a wide range of biological processes
  including cell proliferation, migration, growth, etc.

\bdm{} provides by design five system properties:
\begin{itemize}
\item \textbf{Agent-based.}
        The \bdm{} project is established to support developmental simulations of
          biological dynamics.
        A good approximation for such in silico simulations is agent-based modeling
          \citep{railsback2019agent}.
        Agents are modeled as discrete objects that perform actions based on
          their current state, behavior, and the surrounding environment.

  \item \textbf{General purpose.}
\bdm{} is developed to become a general-purpose tool for agent-based
          simulation.
To simulate models from various fields, \bdm{}'s software
          design is extensible and modular.

  \item \textbf{Large scale.}
        Biological systems contain a large number of agents.
        The cerebral cortex, for example, comprises approximately 16 billion neurons
          \citep{azevedo_equal_2009}.
        Biologists should not be limited by the number of agents within a simulation.
        Consequently, \bdm{} is designed to take full advantage of modern hardware and use
          memory efficiently to scale up simulations.

  \item \textbf{Easily programmable.
        }
The success of a simulator depends, among other things, on how quickly a
          scientist, not necessarily an expert in computer science or high-performance programming,
          can translate an idea into a simulation.
        This characteristic can be broken down into four key requirements that \bdm{} is designed to fullfil:

        First, \bdm{} provides a wide range of common functionalities such as visualization, plotting,
          parameter parsing, backups, etc.
        Second, \bdm{} provides simulation primitives that minimize the
          programming effort necessary to build a use case.
        Third, as outlined in item ``General purpose", \bdm{} has a modular and
          extensible design.
Fourth, \bdm{} provides a coherent API and hides implementation details
          that are irrelevant for a computational model (e.g., details such as parallelization strategy,
          synchronization, load balancing, or hardware optimizations).

  \item \textbf{Quality assured.}
        \bdm{} establishes a rigorous, test-driven development process to foster
          correctness, maintainability of the codebase, and reproducibility of results.

\end{itemize}

The main contribution of this paper is an open-source, high-performance, and general-purpose 
  simulation platform for agent-based simulations.
We provide the following evidence to support this claim:
(i) We detail the user-facing features of \bdm{} that enable users to build
  a simulation based on predefined building blocks and to define a model
  tailored to their needs.
(ii) We present three basic use cases in the field of neuroscience, oncology, and
  epidemiology to demonstrate \bdm{}'s capabilities and modularity.
(iii) We show that \bdm{} can produce biologically-meaningful simulation results
  by validating these use cases against experimental data, or an analytical solution.
(iv) We present performance data on different systems and scale each use case to 
  one billion agents to demonstrate \bdm{}'s performance.

\subsection{Prior work}
\label{sec:prior-work}

The history of agent-based modeling and simulation goes well before the 1990s; 
however, it has seen widespread use in biological systems in the 2000s. 
Several simulators have been published demonstrating the importance of agent-based models in computational biology research
\citep{tisue2004netlogo, emonet_agentcell:_2005, 
    ZublerDouglas2009framework, koene_netmorph:_2009, 
    richmond_high_2010, collier2011repasthpc, lardon_idynomics:_2011, 
    rudge_computational_2012, mirams_chaste:_2013,
    torben-nielsen_context-aware_2014, kang_biocellion:_2014, 
    cytowski_large-scale_2014,
    matyjaszkiewicz_bsim_2017,
    ghaffarizadeh_physicell:_2018}.
In this section, we compare \bdm{}'s most crucial system properties with prior work.

\textbf{Large-scale model support.}
The authors of BioCellion \citep{kang_biocellion:_2014}, PhysiCell
  \citep{ghaffarizadeh_physicell:_2018}, Timothy \citep{cytowski_large-scale_2014},
  and Repast HPC \citep{collier2011repasthpc}
  recognize the necessity for efficient
  implementations to enable large-scale models.
Although these tools can simulate a large number of agents, they do
  not support neural development.
The NeuroMaC neuroscientific simulator \citep{torben-nielsen_context-aware_2014}
  claims to be scalable, but the authors do not present performance data and
  present simulations with only 100 neurons.
Therefore, \bdm{}'s ability to simulate large-scale neural development,
  which we demonstrate in the results section, is, to our knowledge, unrivaled.

\textbf{General-purpose platform.}
Many simulators focus on a specific application area: bacterial colonies
  \citep{emonet_agentcell:_2005, matyjaszkiewicz_bsim_2017,
    rudge_computational_2012, lardon_idynomics:_2011}, cell colonies
  \citep{kang_biocellion:_2014, mirams_chaste:_2013, cytowski_large-scale_2014},
  and neural development \citep{ZublerDouglas2009framework, koene_netmorph:_2009,
    torben-nielsen_context-aware_2014}.
Pronounced specialization of a simulator may prevent its capacity to adapt to 
  different use cases or simulation scenarios.
In contrast, \bdm{} is a general-purpose platform for agent-based simulations by being both modular and extensible.

\textbf{Quality assurance.}
Automated software testing is the foundation of a modern development workflow.
Unfortunately, several simulation tools
  \citep{ZublerDouglas2009framework, rudge_computational_2012,
  lardon_idynomics:_2011, koene_netmorph:_2009, torben-nielsen_context-aware_2014,
  cytowski_large-scale_2014}
  omit these tests.
\cite{mirams_chaste:_2013} recognize this shortcoming and
  describe a rigorous development workflow in their paper.
\bdm{} has over 280 automated tests which are continuously executed on all supported operating systems to ensure high code quality. 
\bdm{}'s open-source code base, tutorials, and documentation not only help users get started, but also enable validation by external examiners.

\section{Design and implementation}

In this section, we present the main simulation concepts of \bdm{} and describe our
  approach to achieve modularity, extensibility, and high performance.
We provide further information about the biological model, software quality, and 
  features like web-based interactive notebooks, and backups in Supplementary File S1 Section~1.

\subsection{Simulation concepts}

\begin{figure}[!htb]
\centering
\includegraphics[width=0.7\linewidth]{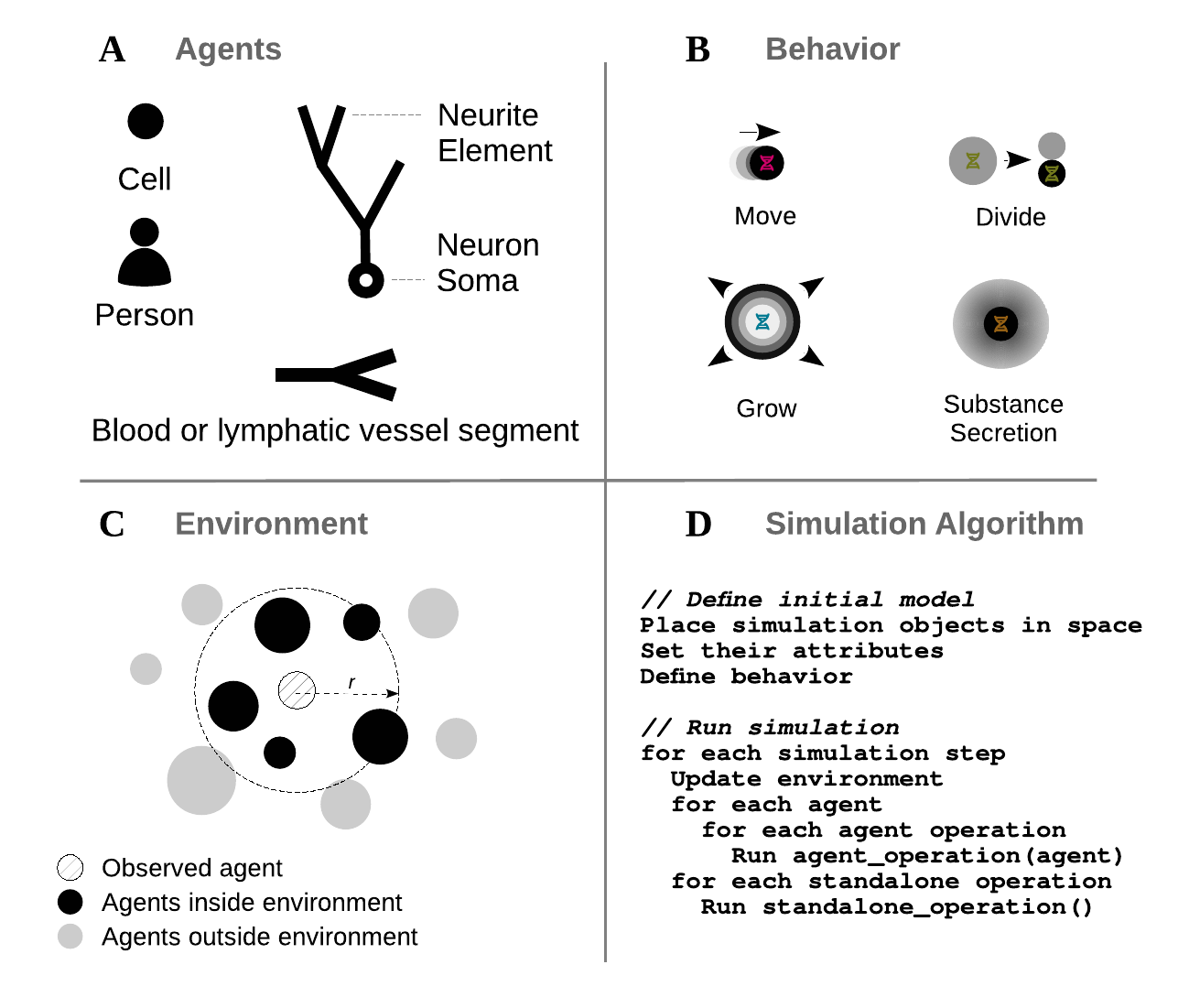}
\caption{{\bf Simulation concepts.}
    Overview of the high-level simulation concepts of \bdm{}. 
    Agents (A) have their own geometry, behavior (B), and environment (C). 
    (B) Agent behavior
    is defined in separate classes, which are inserted into agents.
    A few possible examples for agents and behaviors are displayed.
The update of an agent
    is based on its current state and its surrounding environment. (C) The
    environment is determined by radius $r$ and contains other agents or extracellular
    substances.
    The simulation algorithm (D) can be divided into two main parts: the definition of the
    initial model and execution of the simulation. 
}
    \label{fig:simulation-concepts}
\end{figure}

\bdm{} is implemented in the C++ programming language and supports simulations that follow an agent-based approach.
Figure~\ref{fig:simulation-concepts} gives an overview of \bdm{}'s main concepts, 
  while Figure~\ref{fig:sw-design} illustrates its object-oriented design.

A characteristic property of agent-based simulations is the absence of a
  central organization unit that orchestrates the behavior of all agents.
Quite to the contrary, each agent is an autonomous entity that
  individually determines its behavior.
An agent (Figure~\ref{fig:simulation-concepts}A) has a 3D geometry, behavior, and environment.
There is a broad spectrum of entities that can be modeled as an agent.
In the results section we show examples where an agent represents a subcellular structure
(neuroscience use case), a cell (oncology use case), or an entire 
person (epidemiology use case).
Currently, \bdm{} supports agents with cylindrical and spherical geometry. 
Figure~\ref{fig:simulation-concepts}B shows example agent behaviors such as growth factor secretion, chemotaxis, or cell division.
Like genes, behaviors can be activated or inhibited.
\bdm{} achieves this by attaching them to or removing them from the corresponding
  agent.
\bdm{} simplifies the regulation of behaviors if new agents are created.
The user can control if a behavior will be copied to a new agent or
  removed from the existing agent, based on the event type.

The \textit{Environment} is the vicinity that the agent can
  interact with (Figure~\ref{fig:simulation-concepts}C).
It comprises other agents and chemical substances in the
  extracellular matrix.
Surrounding agents are, for example, needed to calculate mechanical
  interactions among agent pairs.
\bdm{} determines the environment
  based on a uniform grid implementation.
The implementation divides the total simulation space into uniform boxes of
  the same size and assigns agents to boxes based on the center of
  mass of the agent.
Hence, the agents in the environment can be obtained by
  iterating over the assigned box and all its surrounding boxes
  (27 boxes in total).
The box size is chosen based on the largest agent in the simulation to
  ensure all mechanical interactions are taken into account.

Currently, the user defines a simulation programmatically in C++ 
  (Figure~\ref{fig:simulation-concepts}D).
There are two main steps involved: initialization and execution.
During initialization, the user places agents in space, sets
  their attributes, and defines their behavior.
In the execution step, the simulation engine evaluates the defined model in the
  simulated physical environment by executing a series of operations.
We distinguish between agent operations and standalone operations (Figure~\ref{fig:sw-design}).
At a high level, an agent operation is a function that: (i)
  alters the state of an agent and potentially also its
  environment, (ii) creates a new agent, or (iii) removes an
  agent from the simulation.
Examples for agent operations are: execute all behaviors and calculate mechanical forces.
The simulation engine executes agent operations for each agent for each
  time step.
Alternatively, standalone operations perform a specific task during one time step and are 
  therefore only invoked once.
Examples include the update of substance diffusion and the export of visualization data.
Supplementary File S1 Section~1.1.3 contains more details about how operations enable multi-scale simulations.

\subsection{Modularity}
\begin{figure*}[!ht]
\centering
\includegraphics[width=\textwidth]{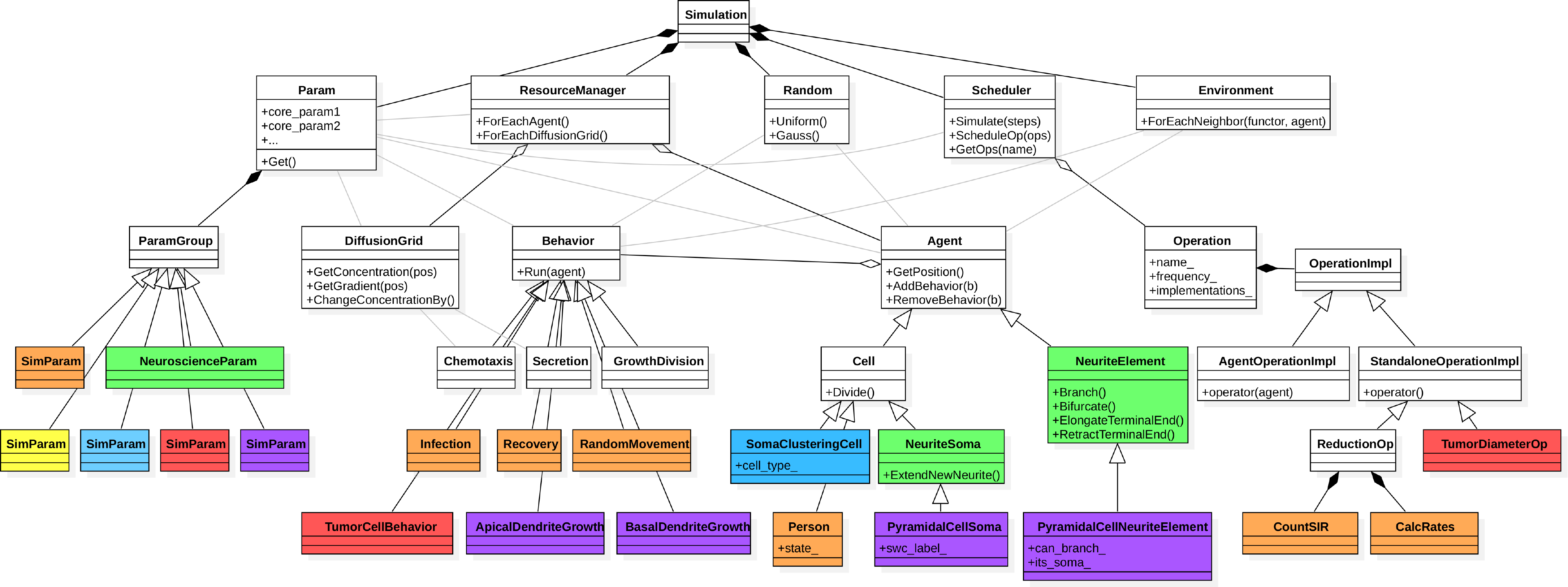}
\caption{{\bf Software design and modularity.}
    Overview of selected classes and functions that are important from the users's perspective.
    Classes in white (\bdm{} core) and green (\bdm{}'s neuroscience module) are part of the current \bdm{} installation.
    The remaining classes illustrate how we extended \bdm{} to implement the use cases and benchmarks shown in this paper
    (purple: neuroscience use case, red: oncology use case, orange: epidemiology use case, blue: soma clustering benchmark, yellow: cell proliferation benchmark).
    A complete list of \bdm{} classes can be found at \href{https://biodynamo.org/}{https://biodynamo.org/}. 
    }
    \label{fig:sw-design}
\end{figure*}

\bdm{} is a simulation platform that can be used to develop
  simulations in various computational biology fields (e.g. neuroscience,
  oncology, epidemiology, etc.).
Although agent-based models in these different fields may intrinsically
  vary, there is a set of functionalities and definitions that they have in
  common.
These commonalities, which consist of
  simulation and support features, are part of the \bdm{} core.
Simulation features include the physics between cellular bodies,
  the diffusion of extracellular substances, and basic behavior, such
  as proliferation and cell death.
Support features include visualization, data analysis, plotting,
  parameter management, simulation backups, etc.
Functionalities that are field-specific are separated from the core and are
  bundled as a separate module.
Figure~\ref{fig:sw-design} gives an overview of \bdm{}'s software design. 
\cite{DEMONTIGNY2020} demonstrated \bdm{}'s modularity by coupling it with another simulator
  to create a hybrid agent-based, continuum-based model.

\textbf{Neuroscience module.}
The neuroscience module is an example of how to extend functionality in the core to
  target \bdm{} to a specific field.
The module adds two new agents \texttt{NeuronSoma} and
  \texttt{NeuriteElement}, and models behavior like neurite extension from
  the soma, neurite elongation, and neurite bifurcation.
The model closely follows the principles of Cortex3D
  \citep{ZublerDouglas2009framework}.
Neurites are implemented as a binary tree.
Each neurite element can have up to two daughter elements.
The cylindrical neurite element is approximated as a spring with a point mass
  on its distal end.
These springs are connected to each other to transmit forces along the chain of
  neurite elements.

\textbf{User-defined components.}
If the desired functionality is missing, the user can create, extend, or modify
  agents, behaviors, operations, and other classes as shown in
  Figure~\ref{fig:sw-design}.
\bdm{}'s software design focuses on loosely-coupled, well-defined components.
This focus not only serves the purpose of creating a clear separation of the
  functionalities of \bdm{}, but, perhaps even more significantly, allows users to integrate
  user-defined components without significant changes to the underlying software
  architecture. This facilitates collaboration and the creation of an open-model library.
We anticipate this library will help researchers in implementing their models more rapidly.

\subsection{Performance and parallelism}

\bdm{}'s performance is based on the following seven enhancements:
(i) The whole simulation engine is parallelized using OpenMP \citep{openmp} compiler
  directives.
OpenMP is a good fit since \bdm{} exploits mostly loop parallelism (see
  Figure~\ref{fig:simulation-concepts}A).
(ii) To increase the maximum theoretical speedup due to parallel processing
  (as described by Amdahl's law
  \citep{amdahl_validity_1967}), we minimize the number of serial code portions in
  \bdm{}.
(iii) We avoid unnecessary copying of data and optimize data access patterns on
  machines with non-uniform memory access (NUMA) architecture.
Compute nodes with multiple NUMA nodes have different memory access latencies
  depending on whether a thread accesses local or remote memory.
Therefore, we load-balance agents and their environment on
  available NUMA nodes.
We built an optimized iterator over all agents to
  minimize threads' memory accesses to non-local memory.
This is necessary because OpenMP does not have built-in support for such
  functionality.
(iv) We detect stationary regions within the simulation and skip the expensive collision
  detection for those agents.
(v) We perform just-in-time compilation to give the visualization engine ParaView direct access 
  to Agent attributes.
(vi) We develop an optimized memory allocator and concurrent hashmap.
(vii) We consider offloading computations to hardware accelerator in our software design 
  (see Figure~\ref{fig:sw-design}).
Our GPU code is implemented in NVidia CUDA 
and OpenCL 
and can
  be executed on graphics cards of different vendors (NVidia, AMD, or Intel).
More details on \bdm{}'s performance enhancements and analyses are beyond this paper's scope,
  and we aim to report them in a future publication.

\section{Results}
This section demonstrates \bdm{}'s capacity to simulate disparate problems in 
  systems biology with simple yet representative use cases in neuroscience, 
  oncology, and epidemiology.
Furthermore, we compare \bdm{}'s performance with an established serial neural simulator
  \citep{ZublerDouglas2009framework}, analyze its scalability, and quantify the impact 
  of GPU acceleration.
For each use case we provide pseudocode for all agent behaviors, a table with 
  model parameters, and more detailed performance results in Supplementary File~S1 Section~2.

\subsection{Neuroscience use case}
\label{sec:pyramidal-cell}

This example illustrates the use of \bdm{} to model neurite growth of pyramidal
  cells using chemical cues.
Initially, a pyramidal cell, composed of a 10 $\mu m$ cell body, three 0.5 $\mu m$ long
  basal dendrites, and one 0.5 $\mu m$ long apical dendrite (all of them considered 
  here as agents), is created in 3D space.
Furthermore, two artificial growth factors were initialized, following a Gaussian
  distribution along the z-axis.
The distribution of these growth factors guided dendrite growth and remained
  unchanged during the simulation.

Dendritic development was dictated by a behavior defining growth
  direction, speed, and branching behavior for apical and basal dendrites.
At each step of the simulation, the dendritic growth direction depended on
  the local chemical growth factor gradient, the dendrite's previous direction,
  and a randomly chosen direction.
In addition, the dendrite's diameter tapered as it grew (shrinkage), until it
  reached a specified diameter, preventing it from growing any further.
The weight of each element on the direction varied between apical and basal
  dendrites.
Apical dendrites were more driven by the chemical gradient and
  were growing at twice the speed of basal dendrites.
On the contrary, basal dendrites were more conservative in their growth
  direction; the weight of their previous direction was more important.
Likewise, branching behavior differed between apical and basal dendrites.
In addition to a higher probability of branching (0.038 and 0.006 for
  apical and basal respectively), apical dendrites had the possibility to
  branch only on the main branch of the arbor.
On the contrary, basal dendrites were only ruled by a simple probability to
  branch at each time step.

\begin{figure}[!h]
  \centering
  \includegraphics[width=0.9\linewidth]{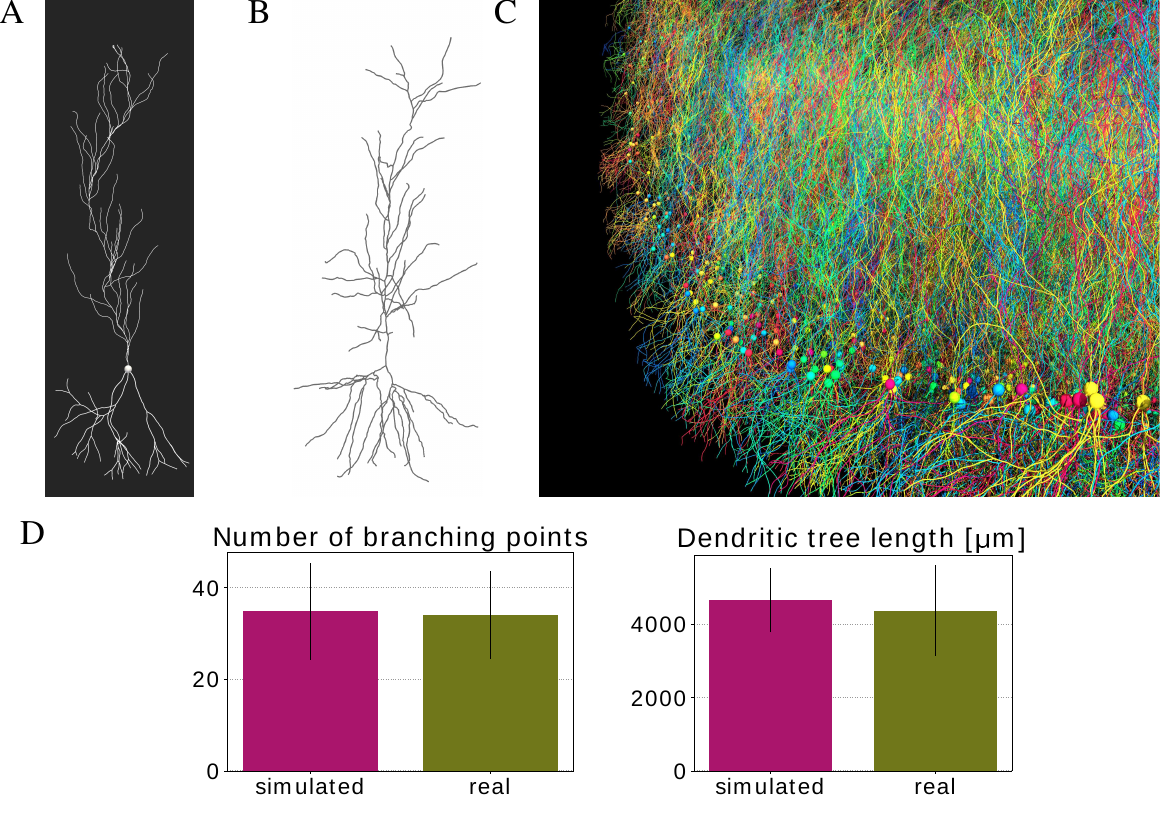}
  \caption{{\bf Pyramidal cell simulation.}
  (A) Example pyramidal cell simulated with \bdm{}.
  (B) Real neuron (R67nr67b-CEL1.CNG) taken from \citep{mellstrom_specific_2016} and 
    visualized with \href{https://neuroinformatics.nl/HBP/morphology-viewer/}{https://neuroinformatics.nl/HBP/morphology-viewer/}
  (C) Large-scale simulation.  
      The model started with 5000 initial pyramidal cell bodies and contained
      9 million agents after simulating 500 iterations.
      Simulation execution time was 46 seconds on a server with 72 CPU cores.
  (D) Morphology comparison between simulated neurons and experimental data from
    \citep{mellstrom_specific_2016}. Error bars represent the standard deviation.
  (A,C) A video is available in Supplementary Information.}
    \label{fig:pyramidal-cell}
\end{figure}

These simple rules gave rise to a straight long apical dendrite with a simple
  branching pattern and more dispersed basal dendrites, as shown in
  Figure~\ref{fig:pyramidal-cell}A, similar to what can be observed in real
  pyramidal cell morphologies as shown in Figure~\ref{fig:pyramidal-cell}B or \cite{spruston2008pyramidal} (Figure~1A CA1).
Using our growth model, we were able to generate a large number of various
  realistic pyramidal cell morphologies.
We used a publicly available database of real pyramidal cells coming from
  \citep{mellstrom_specific_2016} for comparison and parameter tuning.
Two measures were used to compare our simulated neurons and the 69 neurons
  composing the real morphologies database: the average number of branching points,
  and the average length of dendritic trees.
No significant differences were observed between our simulated neurons and the
  real ones ($p < 0.001$ using a T-test for two independent samples).
These results are shown in Figure~\ref{fig:pyramidal-cell}D.
The simulation of the pyramidal cell growth consisted of 361 lines of C++ code.

Figure~\ref{fig:pyramidal-cell}C shows a large scale simulation incorporating
  5000 neurons similar to the one described above, and demonstrates the potential of
  \bdm{} for developmental, anatomical, and connectivity studies in the brain.
This simulation contained 9 million agents.
These 500 iterations correspond to approximately three weeks of pyramidal cell
  growth in the rat.

\subsection{Oncology use case}

In this section, we present a tumor spheroid simulation to replicate
  in vitro experiments from \citep{Gongetal2015invitroMCF7}.
Tumor spheroid experiments are typically employed to investigate the
  pathophysiology of cancer, and are also being used for pre-clinical drug screening
  \citep{Nunes_et_al:2019}.
Here we considered three in vitro test cases using a breast adenocarcinoma
  MCF-7 cell line \citep{Gongetal2015invitroMCF7} with different initial cell
  populations ($2000$, $4000$, and $8000$ MCF-7 cells).
Our goal was to simulate the growth of this mono cell culture embedded in a
  collagenous (extracellular) matrix.
This approach, as opposed to a free suspension one, incorporates cell-matrix interactions
  to mimic the tumor-host environment.

Initially, cancer cells (agents) were clustered in a spherical shape around the origin with a
  diameter of $310$, $380$, or $460$ micrometers.
The three-dimensional extracellular matrix (ECM) was represented in
  our simulations as a $8$ mm\textsuperscript{3} cube.
The fundamental cellular mechanisms modeled here include cell growth,
  cell duplication, cell migration, and cell apoptosis.
A single behavior governed all these processes.
The cell growth rate was derived from the published data \citep{Sutherland3998}, while cell migration
  (cell movement speed), cell survival, and apoptosis were fine-tuned after
  trial and error testing.
Since the in vitro study considered the same agarose gel matrix composition among the experiments,
  the \bdm{} model assumes identical parameters for the cell--matrix interactions in the simulations.
Considering the homogeneous ECM properties, tumor cell migration was
  modeled as Brownian motion.

\begin{figure}[!t]
  \centering
\includegraphics[width=0.9\linewidth]{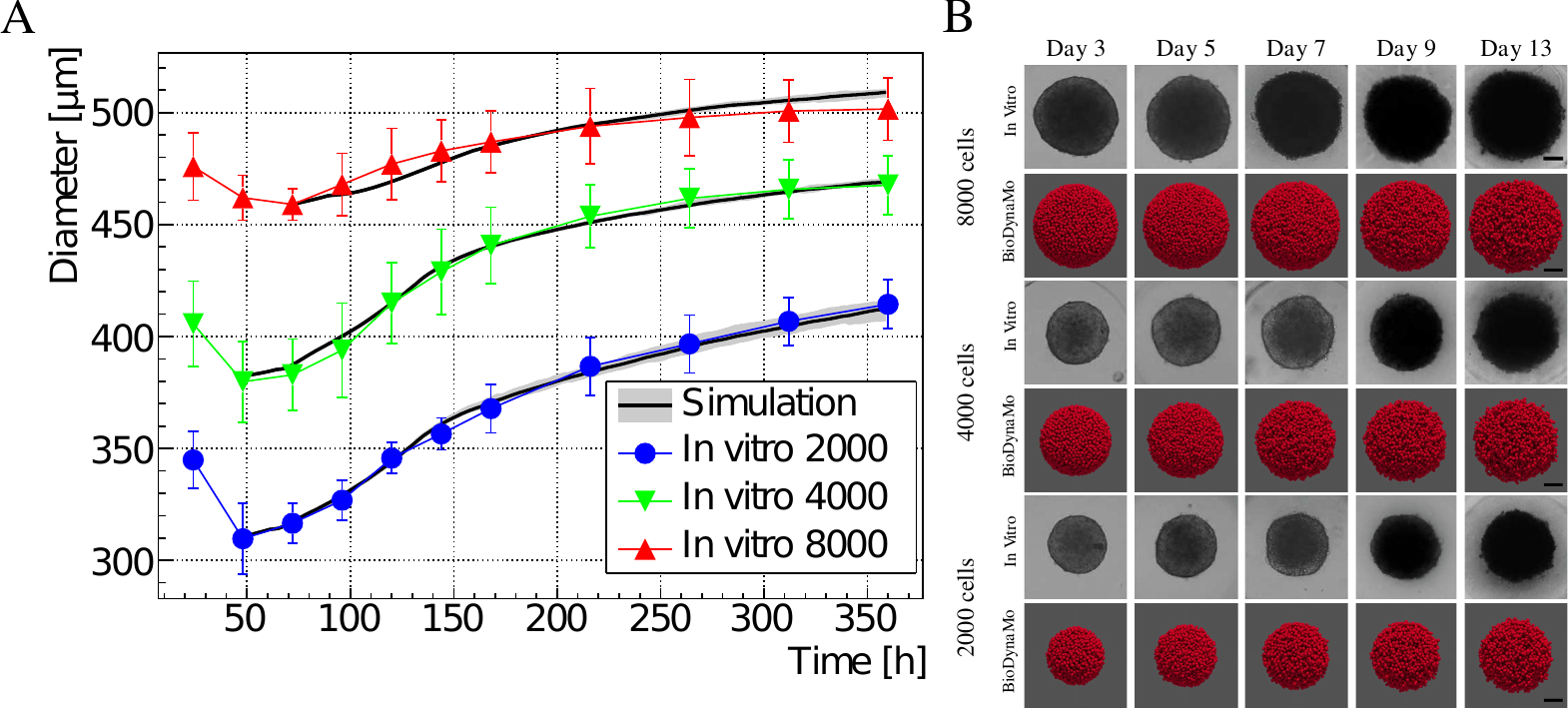}
\caption{{\bf Comparison between in vitro
        MCF-7 tumor spheroid experiments and our in silico simulations using
          BioDynaMo.
      }
    (A) Human breast adenocarcinoma tumor spheroid (MCF-7 cell line) development
      during a 15 day period, where different initial cell populations were
      considered (see Fig 3 in \citep{Gongetal2015invitroMCF7}).
    Error bars denote standard deviation to the experimental data.
    The mean of the in silico results is shown as a solid black line with
    a grey band depicting minimum and maximum observed value.
    (B) Qualitative comparison between the microscopy images and simulation
      snap-shots is shown in the three boxes.
    Scale bars correspond to 100$\mu$m.
    A video is available in Supplementary Information.
  }
  \label{fig:tumor-spheroids}
\end{figure}

The in vitro experiments showed that instantaneous spheroid growth was
  hindered by the compression of the surrounding agarose gel matrix (see
  Figure~\ref{fig:tumor-spheroids}A), owing to cell reorganization at the onset of the cancer mass implantation into the gel.
As a result, the tumor spheroid diameter was initially decreasing.
However, the present simulation example focuses modeling the growth of the spheroid after it had set in the agarose gel matrix.
Therefore, as shown in Figure~\ref{fig:tumor-spheroids}A, \bdm{} simulations are set to start on day two or three.

The in vitro experiments from
  \citep{Gongetal2015invitroMCF7} and the simulations using
  \bdm{} are depicted in Figure~\ref{fig:tumor-spheroids}.
Each line plot in Figure~\ref{fig:tumor-spheroids}A compares the mean diameter
  between the experiments and the simulations
  over time, which demonstrates the validity and accuracy of \bdm{}.
The diameter of the spheroids in the simulations were deducted from the volume
  of the convex hull that enclosed all cancer cells.
The in vitro experiments used microscopy imaging to measure the spheroid's
  diameters \citep{Gongetal2015invitroMCF7}.
Figure~\ref{fig:tumor-spheroids}B compares snapshots of the
  simulated tumor spheroids (bottom row) against microscopy images of in vitro
  spheroids (top row) at different time points.
The spheroid's morphologies between the in vitro experiments and the
  \bdm{} simulations are in excellent agreement.

The example has 424 lines of C++ code, including the generation of the plot shown
  in Figure~\ref{fig:tumor-spheroids}A.
Running one simulation took 0.98--3.39s on a laptop
  and 1.24--4.16s on a server, both using one CPU core.

\subsection{Epidemiology use case}

This section presents an agent-based model that describes the spreading of infectious diseases between humans.
The model divides the population into three groups: susceptible, infected, and recovered 
  (SIR) agents.
We compare our simulation results with the solution of the original SIR model from 
  \cite{kermack_1927}, which used the following three differential equation to describe the model dynamics:
$dS/dt = - \beta S I / N$, 
$dI/dt = \beta S I / N - \gamma I$, and
$dR/dt = \gamma I$.
$S$, $I$, and $R$ are the number of susceptible, infected, and recovered individuals, $N$ is the
  total number of individuals, $\beta$ is the mean transmission rate, and $\gamma$ the recovery rate.

\begin{figure}
  \centering
\includegraphics[width=0.95\linewidth]{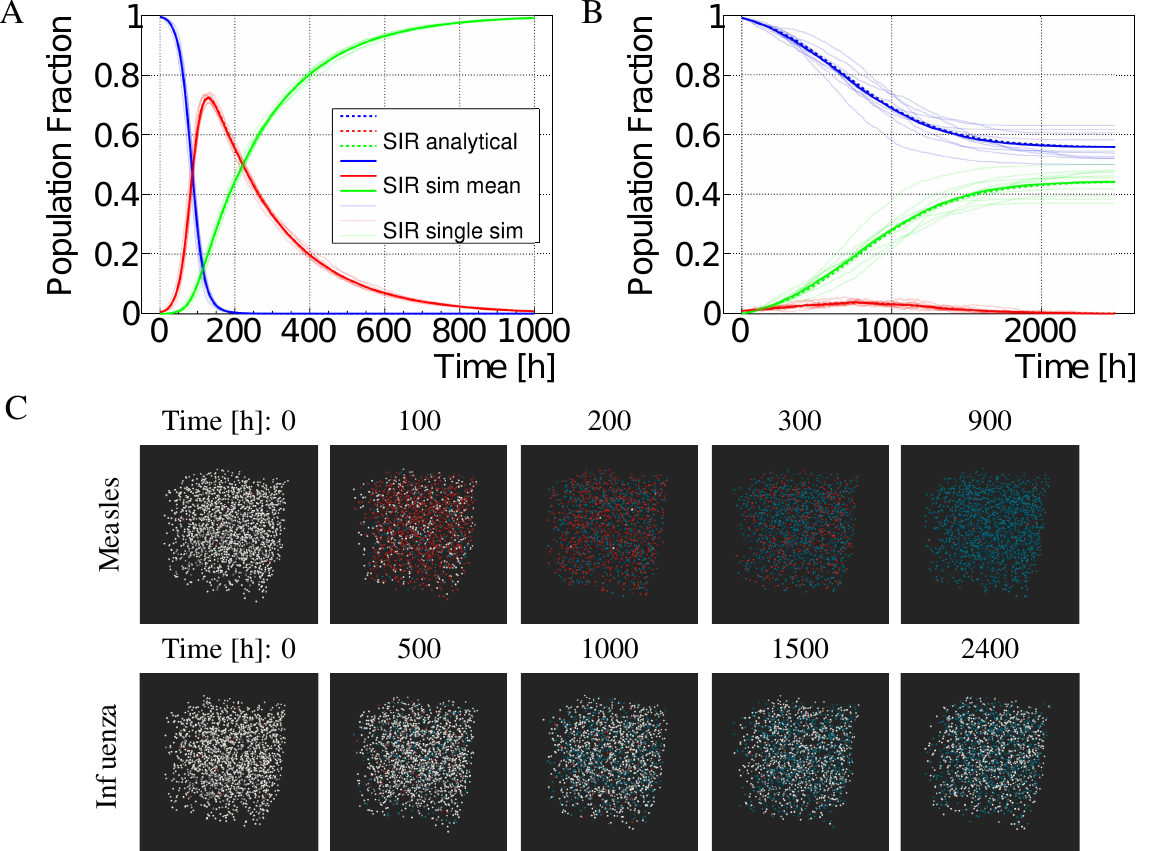}
\caption{{\bf Measles and seasonal influenza SIR model results.}
    (A,B) Comparison between agent-based (solid lines) and analytical (dashed lines) model for
    measles (A) and seasonal influenza (B). The agent-based simulation was repeated ten times. The 
    individual simulation results are shown as thin solid lines. The bold solid line represents the mean
    from all simulations. The legend is shared between the two plots.
    (C) Visualization of the measles and influenza model for different time steps in 3D space.
    Susceptible persons are shown in white, infected persons in red, and recovered persons in blue.
    Persons move randomly and follow the rules for infection and recovery.
}
\label{fig:epidemiology}
\end{figure}

For our agent-based implementation (Figure~\ref{fig:epidemiology}C) we created a new agent
  (representing a person) that encompasses three new behaviors, and extended an operation 
  to count the number of agents in each group (see Figure~\ref{fig:sw-design}).
Agents were randomly distributed in space and have three behaviors.
Infection. A susceptible agent became infected with the infection probability if an infected agent was within the infection radius.
Recovery. An infected agent recovered with the recovery probability at every time step.
Random movement. All agents moved randomly in space. The absolute distance an agent may travel in every time step is limited.

In this agent-based model, the speed at which an infectious disease spreads depended on: the infection probability, the number of contacts each agent has 
  with other agents, and the recovery rate.
The number of contacts in turn depended on the infection radius, the maximum distance an agent may travel, and the density of agents in the simulation space.

We selected two infectious diseases with different characteristics to verify our model: measles and seasonal influenza.
We obtained values for the basic reproduction number $R_0$ and recovery duration $T_R$ from the literature  
(Measles: $R_0 = 12.9$, $T_R = 8$ days \citep{guerra_2017, who_measles}, Influenza: 
  $R_0 = 1.3$, $T_R = 4.1$ days \citep{chowell_2008})
  and determined the parameters $\beta$ and $\gamma$ for the analytical model, based on $R_0 = \beta / \gamma$  and $\gamma = 1 / T_R$.  
For the agent-based model we set the recovery probability to $\gamma$, and placed 2000 susceptible agents and a few infected agents randomly in a cubic space with length 100.
The remaining parameters (infection radius, infection probability, and maximum movement in one time step) were determined using particle swarm optimization \citep{kennedy_1995}.
Figure~\ref{fig:epidemiology} shows that the agent-based model is in excellent agreement with the equation-based approach from \citep{kermack_1927} for measles and influenza.

The example has 566 lines of C++ code, including the generation of the plot shown
  in Figure~\ref{fig:epidemiology}.
Running one simulation took 0.59--1.59s using one CPU core.

\subsection{Performance}

Efficient usage of computing resources is paramount for large-scale simulations with billions of agents, reduced computational costs, and low energy footprint.
To this end, we quantify the performance of \bdm{} with three simulations: cell
  growth and division, soma clustering, and pyramidal cell growth.
These simulations have different properties and are, therefore, well suited to
   evaluate \bdm{}'s simulation engine under a broad set of conditions.
Supplementary File S1 Section~2.2 contains more details about these benchmarks.

\begin{figure}[!t]
\includegraphics[width=\linewidth]{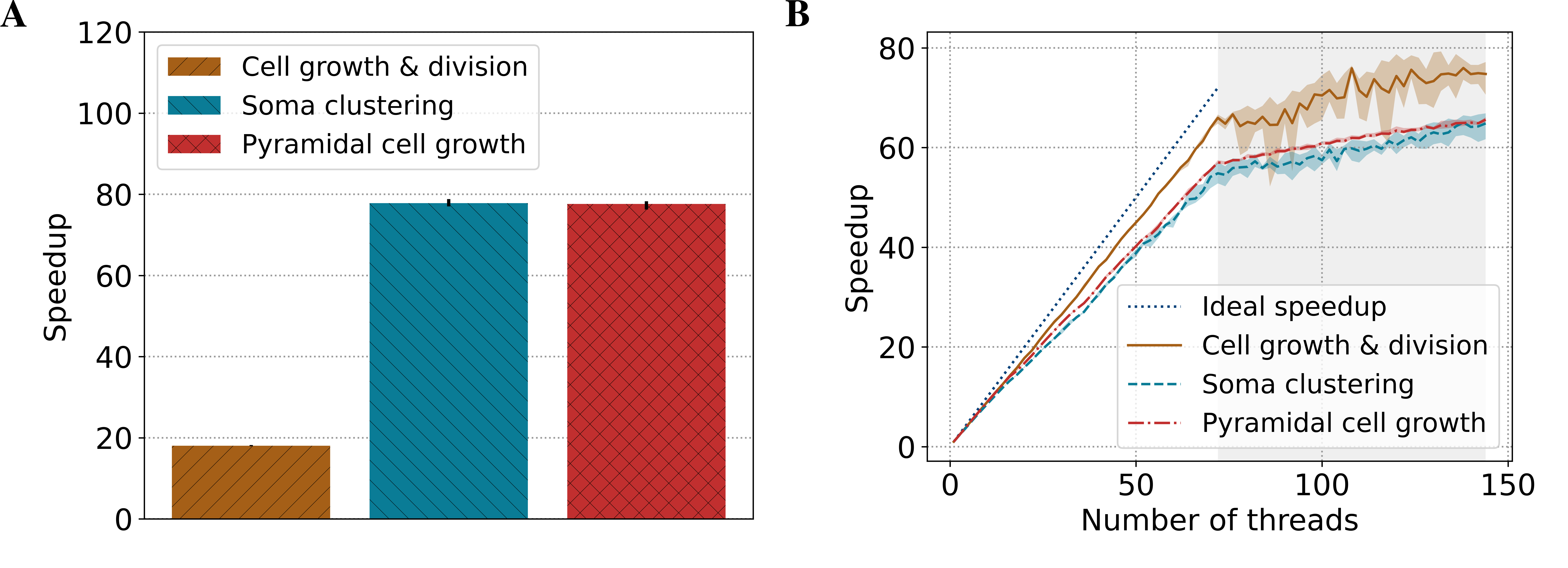}

  \caption{{\bf \bdm{} performance analysis.}
    (A) Speedup of \bdm{} compared to Cortex3D.
    (B) Strong scaling behavior of \bdm{} on a server with 72
      physical cores, two threads per core, and four NUMA domains.
    The grey area highlights hyper-threads.
  }
  \label{fig:performance}
\end{figure}

First, to demonstrate the performance improvements against established agent-based 
  simulators, we compared \bdm{} with Cortex3D \citep{ZublerDouglas2009framework}.
Cortex3D has the highest similarity in terms of the underlying biological model
  out of all the related works presented in Section~\ref{sec:prior-work}.
More specifically, \bdm{} and Cortex3D use the same method to determine mechanical forces
 between agents and the same model to grow neural morphologies. 
This makes Cortex3D the best candidate with which to compare \bdm{}
  and ensure a fair comparison.
Figure~\ref{fig:performance}A shows the speedup of \bdm{} for the three
  simulations.
We observed a significant speedup between 18 and 78$\times$.
Note that we set the number of threads available to \bdm{} to one since
  Cortex3D is not parallelized.
The speedup was larger, when the simulation was more dynamic or more complex.

Second, to evaluate the scalability of \bdm{}, we measured the simulation
  time with an increasing number of threads.
We increased the number of agents used in the comparison with Cortex3D and
  reduced the number of simulation timesteps to 10.
Figure~\ref{fig:performance}B shows the strong scaling analysis.
All simulation parameters were kept constant, and the number of threads was
  increased from one to the number of logical cores provided by the benchmark
  server.
The maximum speedup ranged between 65$\times$ and 75$\times$, which corresponds to a parallel
  efficiency of 0.90 and 1.04.
Performance improved even after all physical cores were utilized and hyper-threads were
  used.
Hyper-threads are highlighted in gray in Figure~\ref{fig:performance}B.
We want to emphasize that even the pyramidal cell growth benchmark scaled well,
  despite the challenges of synchronization and load imbalance.

Third, we evaluated the impact of calculating the mechanical forces on the
  GPU using the cell growth and  division, and soma clustering simulations.
We excluded the pyramidal cell growth simulation because the current GPU kernel
  does not support cylinder geometry yet.
The benchmarks were executed on \systemC, comparing an NVidia Tesla V100 GPU
  with 32 CPU cores (64 threads).
We observed a speedup of 1.27$\times$ for cell growth and division, and
  5.04$\times$ for soma clustering.
The speedup correlated with the number of collisions in the simulation.
The computational intensity is directly linked with the number of collisions
  between agents.

In summary, in the scalability test, we observed a minimum speedup of 65$\times$.
Furthermore, we measured a minimum speedup of 18$\times$ comparing \bdm{}
  with Cortex3D both using a single thread.
Based on these two observations, we conclude that on \systemA{} \bdm{} is
  more than three orders of magnitude faster than Cortex3D.

Based on these speedups, we executed the neuroscience, oncology, and epidemiology use cases with 
  one billion agents. 
Using all 72 physical CPUs on \systemB, we measured a runtime of 1 hour 37 minutes, 
  6 hours 49 minutes, and 3 hours 54 minutes, respectively. 
One billion agents, however, are not the limit. 
The maximum depends on the available memory and accepted execution duration. 
To be consistent across all use cases and keep our pipeline's total execution 
  time better manageable, we decided to run these benchmarks with one billion agents.
Table~5 in Supplementary File S1 shows that available memory would permit an 
  epidemiological simulation with three billion agents. 
With enough memory, \bdm{} is capable of supporting hundreds of billions of agents.

\section{Discussion}

This paper presented \bdm{}, a novel open-source platform for agent-based
  simulations.
Its modular software architecture allows researchers to implement models of distinctly different fields, of which neuroscience, oncology, and epidemiology were demonstrated in this paper.
Although the implemented models follow a simplistic set of rules, the results that emerge from the simulations are prominent and highlight \bdm{}'s capabilities.
We do not claim that these models are novel, but we rather want to emphasize that \bdm{} enables scientists to (i) develop models in various computational biology fields in a modular fashion, (ii) obtain results rapidly with the parallelized execution engine, (iii) scale up the model to billions of agents on a single server, and (iv) produce results that are in agreement with validated experimental data.
Although \bdm{} is modular, we currently offer a limited number of ready-to-use simulation primitives.
We are currently expanding our library of agents and behaviors to facilitate model development beyond the current capacity.

Ongoing work uses \bdm{} to gain insights into retinal development, cryopreservation, 
multiscale (organ-to-cell) cancer modelling, COVID-19 spreading in closed environments, radiation-induced tissue damage, and more.
Further efforts focus on accelerating drug development by replacing in vitro experiments with in silico simulations using \bdm{}.

Our performance analysis showed improvements of up to three orders of magnitude over
  state-of-the-art baseline simulation software, allowing us to scale up simulations to an unprecedented number of agents.
To the best of our knowledge, \bdm{} is the first scalable simulator of 
  neural development with cellular interactions that scales to more than one billion agents.
The same principles used to model axons and dendrites in the neuroscience use case could 
  also be applied to simulate blood and lymphatic vessels.

We envision \bdm{} to become a valuable tool in computational biology, fostering faster and easier simulation of complex and large-scale systems,
  interdisciplinary collaboration, and scientific reproducibility.

\section*{Funding}
This work was supported by the CERN Knowledge Transfer office [to L.B. and A.H.];
the Israeli Innovation Authority [to A.H.];
the Research Excellence Academy from the Faculty of Medical Science of the Newcastle University [to J.dM.];
the UCY StartUp Grant scheme [to V.V.];
the Medical Research Council of the United Kingdom [MR/N015037/1 to R.B., MR/T004347/1 to M.K.];
the Engineering and Physical Sciences Research Council of the UK [EP/S001433/1 to R.B., NS/A000026/1, EP/N031962/1 to M.K.];
a PhD studentship funded by Newcastle University’s School of Computing [to J.J.];
the Wellcome Trust [102037 to M.K.];
the Guangci Professorship Program of Ruijin Hospital (Shanghai Jiao Tong Univ.) [to M.K.];
and by several donations by SAFARI Research Group's industrial partners including Huawei, Intel, Microsoft, and VMware [to O.M.].
The authors have declared that no competing interests exist.

\section*{Acknowledgments}
We want to thank Giovanni De Toni for his work on the \bdm{} build system.

\bibliographystyle{natbib}

\end{document}